\documentclass [english,aps,prb,twocolumn,amsmath,amssymb,showpacs,showkeys]{revtex4}
\usepackage[T1]{fontenc}
\usepackage[latin1]{inputenc}
\usepackage{graphicx}
\usepackage{gensymb}
\makeatletter
\usepackage{epsfig}
\usepackage{prettyref}
\usepackage{babel}
\usepackage{subfigure}
\makeatother
\begin{document}
%%\title{Memory and Aging in NiO nanoparticles}
\title{Memory, Aging and Spin Glass Nature: A Study of NiO Nanoparticles}
\author{Vijay Bisht}
\email{vijayb@iitk.ac.in}
\author{K.P.Rajeev}
\email{kpraj@iitk.ac.in}
 \affiliation {Department of Physics,
Indian Institute of Technology Kanpur 208016, India}

\begin{abstract}
We report studies on magnetization dynamics in NiO nanoparticles of average size 5~nm. Temperature and time dependence of dc
magnetization, wait time dependence of magnetic relaxation (aging) and memory phenomena in the dc magnetization are studied
with various temperature and field protocols. We observe that the system shows memory and aging in field cooled and zero field
cooled magnetization measurements. These experiments show that the magnetic behavior of NiO nanoparticles is similar to spin glasses. We argue that the spin glass behavior originates from the freezing of spins at the surface of the individual particles.
\end{abstract}
\pacs{75.50.Tt, 75.50.Lk, 75.30.Cr, 75.40.Gb}

 \keywords{NiO nanoparticles, magnetic relaxation, aging, memory effects.}

 \maketitle

\section{INTRODUCTION}

The slow dynamics shown by magnetic nanoparticles has been an active area of research for the past two decades because of numerous technological applications as well as for understanding the physics behind the exotic phenomena observed.\cite{Dormann}
 Ferro and ferrimagnetic nanoparticles have been studied more than antiferromagnetic nanoparticles because of their technological
potential as they have high magnetic moments.\cite{Steen} Antiferromagnetic materials show a drastic change in their magnetic properties when the particle size goes to the nano regime because of the uncompensated spins at the surface which
give rise to a net magnetic moment. This leads to many interesting magnetic properties e.g. a bifurcation between field cooled (FC) and zero field cooled (ZFC) magnetization, a peak in ZFC magnetization, slow relaxation of magnetization, wait time
dependence of magnetization relaxation (aging) and memory in FC and ZFC magnetization measurements.\cite{Batlle,Suzuki,Sasaki,Sun,Malay,Chakraverty,Tsoi,RZheng,Martinez} If the particles are non interacting, the magnetization dynamics is described by
superparamagnetic relaxation as predicted by Néel-Brown theory.\cite{Neel,Brown} On the other hand, interactions can give rise to a spin glass like behavior (superspin glass) in interacting nanoparticles.\cite{Batlle,Sasaki,Sun,Malay,Sahoo} However, spin glass behavior can also arise in the nanoparticles due to spin frustration at the
surface of individual particles.\cite{Martinez,Kodama,Tiwari,Winkler}

 Bulk Nickel oxide (NiO) is known to be  antiferromagnetic with a Néel temperature $T_{\textnormal{N}}$ of  523~K. The temperature
dependence of magnetization of NiO nanoparticles was first studied in $1956$ by Richardson and Milligan and a peak in the
magnetic susceptibility was found much below the bulk $T_{\textnormal{N}}$.\cite{Richardson} It was observed that on decreasing the particle size, the magnetization increases
 and the peak in susceptibility shifts to lower temperatures. Later in $1961$ Néel suggested that small antiferromagnetic particles should exhibit superparamagnetism and weak ferromagnetism.\cite{Neel1} The observed particle
moment of NiO nanoparticles is found to be much larger than that predicted by the two lattice model of antiferromagnets and  a multi sublattice model has been proposed to explain it and also the observed high coercivities and loop shifts in
these particles.\cite{Makhlouf,Kodama1} There have been some reports on the magnetic properties of NiO nanoparticles which claim that
they are superparamagnetic.\cite{Ghosh,Jongnam,Richardsona,Fatemeh,MSeehra,Yuko} However, there are issues in considering them as superparamagnetic as their magnetization cannot be described by the modified Langevin function.\cite{Makhlouf} Tiwari et al. have done a detailed study on the magnetic properties of NiO nanoparticles and have claimed, on the basis of scaling arguments, that NiO nanoparticles show spin glass behavior.\cite{Tiwari} They have proposed that the surface spin disorder and
frustration give rise to such behavior. Winkler et al. have done
magnetic measurements on both bare and polymer dispersed NiO nanoparticles of size 3~nm and have found that they can
be thought  to be consisting of an antiferromagnetic core with an uncompensated moment and a disordered surface shell.\cite{Winkler} They have  proposed that the interparticle interactions can increase the effective anisotropy energy of the core magnetic moments which results
in shifting the freezing temperatures to higher values and in  enhancing the frustration of the spins at the
surface. The behavior of NiO nanoparticles is also found to depend on the method of preparation, whether they are coated or not, and the nature of the coating.\cite{Ghosh,Bodker,Seehra,MSeehra,Shim,Pishko, Winkler}

Aging and memory effects have been investigated in many
nanoparticle systems using ac susceptibility and low field dc
magnetization measurements with various temperature and field
protocols.
\cite{Sun, Sasaki, Michael, Sahoo, Raj, Gufei, Zheng,Parker,DParker,Malay,Chakraverty,Tsoi}
Non-interacting particles  are expected to show aging and memory
effects only in FC magnetization measurements. These effects have
been observed by various authors and their explanation  is based
on a simple superparamagnetic model where one assumes a
distribution of anisotropy energy barriers and temperature driven
dynamics.\cite{Sasaki, Zheng, Tsoi, Chakraverty} By contrast, in
interacting particles, the magnetization dynamics is spin glass
like and so it is expected that they would show aging and memory
effects in both FC and ZFC protocols like spin glasses. Indeed,
this is the case and models based on canonical spin glasses have
been used to explain these effects in such nanoparticle
systems.\cite{Sun, Sasaki} Thus the presence of aging and memory in
ZFC protocol is like a litmus test for differentiating spin
glasses and superparamagnets.

Most of the  nanoparticles studied for aging and memory effects
are ferro or ferrimagnetic and there are very few studies on
antiferromagnetic nanoparticles. We feel that it would be
interesting to study these effects in  NiO nanoparticles, an
antiferromagnetic system in which surface effects are known to
play a major role in determining the magnetic behavior. In fact, it has been claimed that these particles show spin glass behavior.\cite{Tiwari, Winkler} In this
work, we present a detailed study on aging and memory effects in
5~nm NiO particles with various temperature and field protocols and try to settle the issue of its spin glass nature.

\section{EXPERIMENTAL DETAILS}

NiO nanoparticles are prepared by the sol gel method.\cite{Makhlouf,Tiwari,Richardsona} Nickel hydroxide precursor is precipitated by
reacting aqueous solutions of nickel nitrate (99.999\%)and sodium hydroxide(99.99\%) at pH~=~12, at room temperature. This
precipitate is washed many times with distilled water to remove remnant nitrate and sodium ions. It is then dried at 100\degree C
for 6 hours to get green colored nickel hydroxide powder. Nickel oxide nanoparticles are prepared by heating nickel hydroxide at
250\degree C for three hours in flowing helium gas. The sample is characterized by X-ray diffraction (XRD) using a Seifert
diffractometer with Cu~K$\alpha$ radiation. The average particle size as determined by XRD using the Scherrer formula is 5~nm.
All the magnetic measurements are done with a SQUID magnetometer (Quantum Design, MPMS XL5).

\section{RESULTS AND DISCUSSION}

\subsection{Aging Experiments}

\begin{figure}[!t]
\begin{centering}
\includegraphics[width=1\columnwidth]{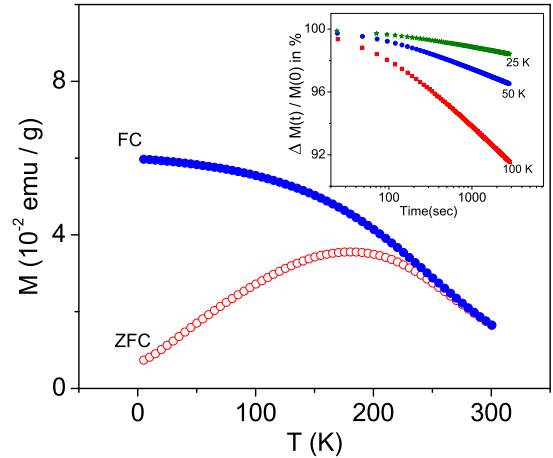}
\par\end{centering}
\caption{(Color online) Temperature dependence of the dc magnetization in a 100~Oe field for both ZFC and FC protocols. The inset
shows decay of thermoremanent magnetization at temperatures 25~K, 50~K and 100~K. } \label{fig:MvsT}
\end{figure}

\begin{figure}[!t]
\begin{centering}
\includegraphics[width=1\columnwidth]{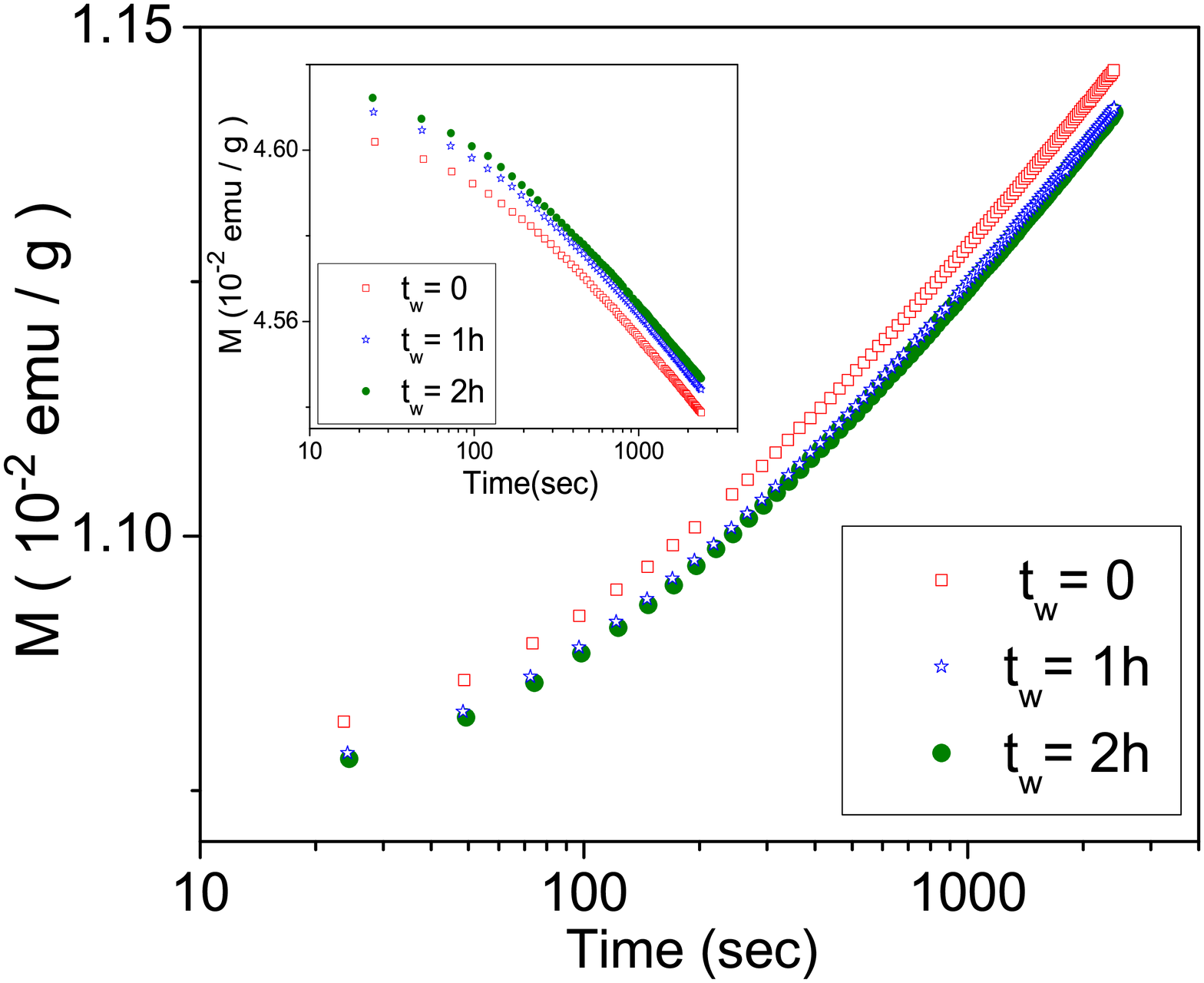}
\par\end{centering}
\caption{(Color online) Wait time dependence of ZFC magnetization
at 25~K. Inset shows the wait time dependence of  TRM at 25~K. }
\label{fig:aging}
\end{figure}

Temperature dependence of magnetization was done under FC and ZFC protocols at a field of 100~Oe. See Figure \ref{fig:MvsT}.
There is a bifurcation in FC and ZFC magnetizations which manifests below 275~K and the ZFC magnetization has a broad peak
at about 180~K. It can be seen that the FC magnetization increases with decreasing temperature apparently tending to saturate.
Time decay of thermoremanent magnetization (TRM) was done at temperatures 25~K, 50~K and 100~K. For these measurements, we cool the
sample in a field of 100~Oe to the temperature of interest and then switch off the field. Now the magnetization is measured as a
function of time. See inset of Figure \ref{fig:MvsT}. It can be observed that the magnetization decays more or less logarithmically. This behavior
is a characteristic of both superparamagnets and spin glasses.
 An experiment that can distinguish between the above two
possibilities is the wait time dependence of magnetization relaxation (aging).
 We carried out aging experiments in both FC and ZFC protocols as
follows: Cool the sample in a field of 100~Oe for FC (or in zero
field for ZFC) to the temperature of interest, wait for a specified time (wait time) and then switch the field off (or on in case of ZFC). Now record the magnetization as a function of time. Superparamagnets are expected to show a weak wait time dependence of TRM  and no wait time dependence in ZFC magnetization; in other words weak FC aging and no ZFC aging. Spin glasses are, however, known to show both FC and ZFC aging.
\cite{Sasaki,Jonsson} Figure~\ref{fig:aging} shows the data for
aging experiments in  FC and ZFC protocols.  A noticeable wait
time dependence in both FC and ZFC protocols  can be observed
which is an evidence in support of spin glass behavior
in NiO nanoparticles.

%Figure 1%

%Figure 2%

%Figure 3%

\begin{figure}[!t]
\begin{centering}
\includegraphics[width=1\columnwidth]{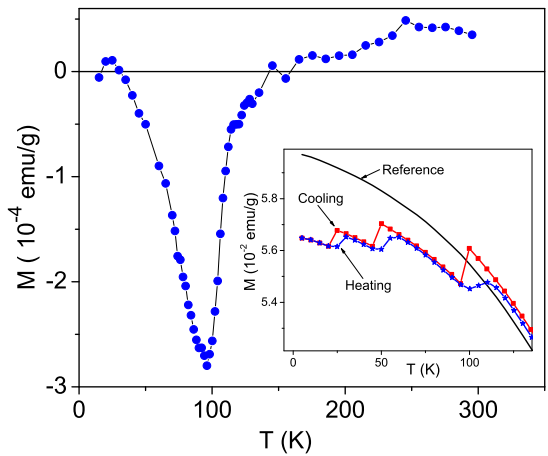}
\par\end{centering}
\caption{(Color online) Memory experiments in ZFC protocol. The
difference in magnetization  with a stop of one hour at 100~K in
the cooling process  and the reference data,
plotted as a function of temperature. Inset: Memory experiments
in FC protocol  with stops of one hour duration at 100~K, 50~K,
and 25~K. The field is switched off during each stop.} \label{fig:ZFC
memory}
\end{figure}

%Figure 4%
\begin{figure}[!b]
\begin{centering}
\includegraphics[width=\linewidth]{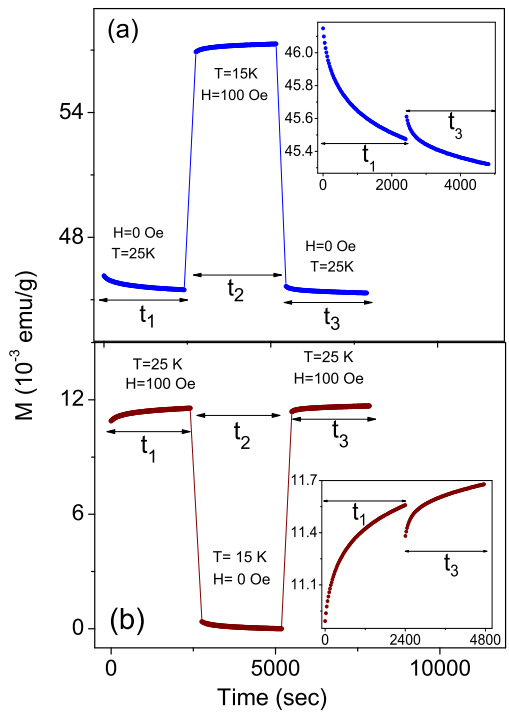}
\par\end{centering}
\caption{(Color online) Magnetic relaxation with negative
temperature cycling and a field change for (a) FC protocol. (b)
ZFC protocol. The insets show that the relaxation  during time
$t_3$ is essentially the continuation of the relaxation during
$t_1$, confirming that the system has the memory of earlier
relaxations.}
\label{fig:Tcycling}
\end{figure}

\subsection{Memory Experiments}

We carried out memory experiments in both FC and ZFC
magnetization measurements. In the ZFC protocol, we first record the ZFC magnetization in the standard way and call this as the reference data. Now the sample is cooled in  zero field to 5~K with a stop of one hour at 100~K. During subsequent heating the magnetization is recorded  up to 300~K. In Figure \ref{fig:ZFC memory}
we show the difference in magnetization between the ZFC  data with the stop and the ZFC reference data. It is clear that there is a dip at 100~K,
where the stop was taken during the cooling process establishing the
ZFC memory in the system. For doing FC memory experiments, the
system is cooled in the presence of a magnetic field to 5~K with
intermittent stops of one hour at 25~K, 50~K and 100~K with the field switched off during the stops. The magnetization is measured while cooling and then during subsequent heating. The data is shown in the inset of Figure
\ref{fig:ZFC memory}. It can be observed that the system remembers
the history of the cooling process and the magnetization takes jumps
close to the temperatures where the stops were taken.

Memory in FC magnetization has  been observed for both interacting and non interacting nanoparticles and it has been shown that a broad distribution of energy barriers is  sufficient to produce memory effects in FC protocol.\cite{Sasaki} However memory in ZFC magnetization is  a  feature
inherent to spin glasses and has not been observed in
superparamagnets. Thus the memory observed in  ZFC magnetization measurements  provides  conclusive evidence in favor of the spin glass  nature of NiO nanoparticles.
However, the width of the dip in Figure \ref{fig:ZFC memory} is rather
large, about 100~K, the corresponding figure for canonical spin glasses being a few Kelvins.\cite{Jonsson1}

To complement these memory experiments we have done negative
temperature cycling experiments with field change in both FC and
ZFC protocols as suggested by Sun et al. and adopted by many
authors.\cite{Sun,Sasaki,RZheng,Tsoi,Malay} In FC protocol, the
system is cooled to 25~K in a field of 100~Oe, the field is
then switched off and the magnetization is recorded for a time period
$t_1$. Then the system is cooled to 15~K, a field of 100~Oe is
applied and magnetization data is taken for a period $t_2$.
Temperature is now changed back to 25~K, field is switched off and
magnetization is recorded again for a period $t_3$. Here $t_1$ = $t_2$ = $t_3$= 2800 seconds. See Figure \ref{fig:Tcycling}(a). It can be
seen that when the temperature is raised back to 25~K, the
relaxation starts almost from the point at which it was left off in the  previous relaxation at 25~K. Please see the inset of Figure \ref{fig:Tcycling}(a). This shows that the
system has a memory of an earlier aging in spite of an
intervening aging at a lower temperature. We have also done negative temperature cycling for ZFC magnetization relaxation in a similar manner. See \ref{fig:Tcycling}(b) and its inset. The results again confirm the  existence of memory in ZFC protocol.

\subsection{Discussion}

The presence of aging and memory in ZFC magnetization of NiO nanoparticles confirms their spin glass behavior. There have been some work on other nanoparticle systems where ZFC memory was observed.\cite{Sun, Sasaki, Raj, Jonsson1, Parker, Suzuki} All those works were on ferri and ferromagnetic materials and the interparticle interactions were said to be responsible for the observed glassy behavior. The dip in the ZFC memory in the present work  (Fig. \ref{fig:ZFC memory}) is quite broad compared to  those reported on other nanoparticle systems. This suggests that the origin of spin glass behavior in NiO nanoparticles is, possibly, not interparticle interactions. In fact, the interactions between these particles are very weak  and are not sufficient to cause collective freezing of particle moments at such high temperatures as has been argued by Tiwari et al.\cite{Tiwari} However these interactions can  enhance the frustration of spins at the surface of individual particles and shift the freezing temperatures to higher values.\cite{Winkler}  The exchange bias effects  observed in NiO nanoparticles indicate  the presence of both ferro and antiferromagnetic interactions at the surface, which can frustrate the spins leading to spin glass behavior.\cite{Yi,Salah}  Thus the origin of spin glass state in NiO  nanoparticles seems to be the  freezing of spins at the surface of the individual particles.  The wide dip in ZFC memory of NiO nanoparticles as compared to canonical spin glasses can  possibly be attributed to the finite size of the system.

 \section{Conclusion}
We have done dc magnetic relaxation measurements on NiO nanoparticles with various temperature and field protocols. Our results show the presence of aging and memory effects in both FC and ZFC magnetizations, thus establishing the spin glass behavior of these particles. The  origin of this behavior seems to be surface spin freezing of individual particles rather than interparticle interactions.

\begin{acknowledgments}
VB thanks the University Grants Commission of India for financial support.
\end{acknowledgments}

\end{document}